\documentclass[a4paper,12pt]{article}
\usepackage[cp866]{inputenc}
\usepackage{hyperref,amsfonts}
\def\be{\begin{equation}}
\def\ee{\end{equation}}
\def\ba{\begin{array}}
\def\ea{\end{array}}
\def\bea{\begin{eqnarray}}
\def\eea{\end{eqnarray}}

\title{\bf Classification of integrable
quadratic Hamiltonians on e(3)}

\author{
Thomas Wolf\\ Department of Mathematics,
Brock University\\ 500 Glenridge Avenue, St.Catharines,\\
Ontario, Canada L2S 3A1\\
twolf@brocku.ca\\ \\ and \\ \\
Olya V. Efimovskaya\\
Department of Mechanics and Mathematics,\\
Moscow State University\\
Vorob'evy gory,\\
Moscow, 119992, Russia\\
efvitaly@ttk.ru
}

\begin{document}
\maketitle

\section{Introduction.} %=========================================
In this work we consider quadratic Hamiltonians of the form
\begin{equation}
H=(M, A M)+(M, B \gamma)+(\gamma, C \gamma)+(P,M)+(Q,\gamma).
\label{HAMGEN}
\end{equation}
where $M=(M_1, M_2, M_3),\ \gamma=(\gamma_1,\gamma_2,\gamma_3)$ and
$A, C$ are constant symmetric matrices, $B$ is a general constant
matrix and $P, Q$ are constant vectors. Without loss of generality we
choose
\begin{equation} A=\mbox{diag}(a_{1},a_{2},a_{3}).\label{diagA}
\end{equation}
The equations of motion are given by
\[ \frac{dM_i}{dt} = \{H,M_i\}, \ \ \frac{d\gamma_i}{dt} = \{H,\gamma_i\}\]
where Poisson brackets are defined by
\begin{equation}
\{M_i,M_j\}=\varepsilon_{ijk}\,M_{k}, \qquad
\{M_{i},\gamma_{j}\}=\varepsilon_{ijk}\,\gamma_{k}, \qquad
\{\gamma_{i},\gamma_{j}\}=0\label{puas}
\end{equation}
with $\varepsilon_{ijk}$ being the totally anti-symmetric tensor.
These linear Poisson brackets are related to the Lie algebra $e(3)$.

Because they admit the two Casimir functions
\begin{equation}
\ J_1 = (\gamma,\gamma), \ \ \ J_2 = (\gamma,M)\label{cas}
\end{equation}
we need only one additional first integral for Liouville integrability
\cite{arn}.  In our paper we restrict ourselves to polynomial first
integrals.  The relevance of the combination of quadratic Hamiltonians
(\ref{HAMGEN}) with Poisson brackets (\ref{puas}) arises from the
Euler-Poinsot model describing motion of a rigid body around a fixed
point under gravity and from the Kirchhoff model describing the motion
of a rigid body in ideal fluid \cite{koz,boma1,boma2}.

In section \ref{known} we give a list of known integrable quadratic
Hamiltonians with additional polynomial integrals of motion.
Results of our search are given in section \ref{results}.

It is known from literature that for pairwise non-equal $a_1,a_2,a_3$
in (\ref{diagA}) no extra integrable cases besides those mentioned in
section \ref{known} exist. In addition all Hamiltonians (\ref{HAMGEN})
with linear or quadratic first integral are known too.  Hamiltonians
with cubic first integral are discussed in \cite{crm}. We therefore
study first integrals $I$ of fourth degree and consider the case
$a_1=a_2\neq a_3$.

The motivation for a systematic investigation of this ansatz arose
from the recent finding of V.\ Sokolov of a new integrable
Hamiltonian of this form with a fourth degree first integral.
Our investigation is in so far more general as we consider
quadratic Hamiltonians {\em with linear} terms. Our complete
classification of this problem resulted in new integrable Hamiltonians
with complex coeffi\-cients. The simplest example of such a kind is
\[H = M_{1}^2 + M_{2}^2 + 2 M_{3}^2 + i M_{1} + M_{2}\]
which commutes with some fourth degree polynomial
under $so(3)$-brackets.

We also were inspired by the Goryachev-Chaplygin Hamiltonian \cite{gor,chap}
which admits a fourth degree integral exclusively on the Casimir level
$J_2=0$. Unfortunately our complete classification did not result in
new integrable cases.

In section \ref{quantum} we follow \cite{kom1,kom2} and consider the quantum
counterpart
\begin{equation} \label{com}
[ M_{i},M_{j}]=\varepsilon_{ijk}\,M_{k}, \qquad
[M_{i},\gamma_{j}]=\varepsilon_{ijk}\,\gamma_{k}, \qquad
[\gamma_{i},\gamma_{j}]=0
\end{equation}
of the Poisson bracket (\ref{puas}). Here $M_i, \gamma_j$ are elements
of an associative algebra with commutator relations (\ref{com}).
The Hamiltonian is a (non-commutative) polynomial of second degree and
first integrals are polynomials which commute with the Hamiltonian.
Also here we classify quadratic Hamiltonians with linear terms having a
fourth degree first integral and obeying the restriction $a_1=a_2\neq a_3$.
We find four cases, among them the quantum analogue
of the above mentioned integrable Hamiltonian of V.\ Sokolov.

In all three investigations an ansatz for the Hamiltonian and an
ansatz for a fourth degree polynomial were made, the Poisson bracket
or commutator were computed and set to zero, resulting in large
non-linear over-determined algebraic systems. Although they are linear in
the coefficients of $H$ and linear in the coefficients of the first
integral, the algebraic systems are still very large (see table 1) and
challenging to solve.

We used the computer algebra program {\sc Crack} which was originally
designed to solve over-determined PDE-systems. But its interactive
capabilities allowed to solve large algebraic systems at first
interactively and in doing so to learn how to take advantage of the
bi-linearity.  Due to an ongoing effort to generalize gathered
experience and to incorporate it into the program it is now able to
solve large bi-linear systems, for example the first and third system
in table 1 automatically and the second system with only few manual
interactions.  An overview of essential features of the program and
other applications requiring the solution of bi-linear systems is
given in \cite{crm}.

\begin{center}
%\scriptsize
\begin{tabular}{|l||c|c|c|} \hline
type of problem   & $\ \ e(3)\ \ $ & $e(3), J_2=0$ & $e(3)$ quant. \\
\hline \hline
\# of unknowns ($H$,$I$,total)
                  & 17,200,217 & 17,176,193 & 17,200,217 \\ \hline
\# of equations   &  451  &  396  & 451  \\ \hline
total \# of terms & 5469  & 5243  & 9681 \\ \hline
average \# of terms/equ.\
                  &  12.1  &  13.2  & 21.5  \\ \hline
time to solve     & 18h 53min & $\approx$ 15h  & 11h 43min  \\ \hline
% solutions        & 12 merged & 16 merged  & 9 merged  \\ \hline
% www address      & e3c2new  & e3nullnew & e3quant  \\ \hline
details in section & \ref{e3c2} & \ref{e3c2null} & \ref{quantum} \\ \hline
\end{tabular} \vspace{6pt}\\
%\normalsize
Table 1. An overview of the solved algebraic systems
\end{center}
Times are measured on a 1.7GHz Pentium 4 running a 120 MByte
REDUCE session under Linux.

\section{Known integrable Hamiltonian on $e(3)$}
\label{known}%==============
In the following we list all known integrable Hamiltonian of type
(\ref{HAMGEN}) on
$e(3)$.
First let us consider all cases where the matrices $A=\{a_{ij}\}$,
$B=\{b_{ij}\}$
and $C=\{c_{ij}\}$ are diagonal,
i.e.\ the Hamiltonian takes the form:
\begin{equation}
\begin{array}{l}
H=a_{1} M_{1}^{2}+a_{2} M_{2}^{2}+a_{3} M_{3}^{2}+ 2 b_{1} M_{1}\gamma_{1}
+2 b_{2}M_{2}\gamma_{2}+2 b_{3} M_{3}\gamma_{3}+\\[3mm]
\quad c_{1}\gamma_{1}^{2}
+c_{2}\gamma_{2}^{2}+c_{3}\gamma_{3}^{2}+p_{1} M_{1}+p_{2} M_{2}+p_{3}
M_{3}+q_{1} \gamma_{1}+q_{2} \gamma_{2}+q_{3} \gamma_{3}.
\label{diagHam}
\end{array}
\end{equation}

{\bf Kirchhoff's problem of the motion of a rigid body in ideal fluid.}
In this case the Hamiltonian does not contain linear terms,
i.e.\ we have $p_{i}=q_{i}=0.$
For this problem there are three known classical integrable cases of
Kirchhoff \cite{Kir}, Clebsch \cite{cleb} and Steklov-Lyapunov
\cite{steklov,lyap}.

The Kirchhoff case is defined by the identities
\[a_1=a_2, \qquad b_{1}=b_{2}, \qquad c_{1}=c_{2}.\]
The additional integral $I$ is linear: $I=M_{3}$.

For the Clebsch case the coefficients $a_i$ are arbitrary and the remaining
parameters satisfy the following conditions:

\[b_{1}=b_{2}=b_{3},\]
\[\frac{c_{1}-c_{2}}{a_3}+\frac{c_{3}-c_{1}}{a_2}+\frac{c_{2}-c_{3}}{a_1}=0.
\]
If not all $a_i$ are equal then the Hamiltonian can be represented in the
form
\[H=a_{1} M_{1}^{2}+a_{2} M_{2}^{2}+a_{3} M_{3}^{2}+a_{2}a_{3}\gamma_{1}^{2}
+a_{3}a_{1}\gamma_{2}^{2}+a_{1}a_{2}\gamma_{3}^{2}.\]
The additional quadratic integral admitted by $H$ is
\[I=M_{1}^{2}+M_{2}^{2}+M_{3}^{2}+(a_{2}+a_{3})\gamma_{1}^{2}
+(a_{3}+a_{1})\gamma_{2}^{2}+(a_{1}+a_{2})\gamma_{3}^{2}.\]
If $a_{1}=a_{2}=a_{3}$ then we get the Neumann Hamiltonian
$$
H=M_{1}^{2}+M_{2}^{2}+M_{3}^{2}+c_{1}\gamma_{1}^{2}
+c_{2}\gamma_{2}^{2}+c_{3}\gamma_{3}^{2}.
$$
The additional integral in this case coincides with $H$.

For the Steklov-Lyapunov case the coefficients $a_i$ are arbitrary and the
remaining parameters satisfy the following conditions:
\[\frac{b_{1}-b_{2}}{a_3}+\frac{b_{3}-b_{1}}{a_2}+\frac{b_{2}-b_{3}}{a_1}=0,
\]
\[c_{1}-\frac{(b_{2}-b_{3})^2}{a_1}=
c_{2}-\frac{(b_{3}-b_{1})^2}{a_2}=
c_{3}-\frac{(b_{1}-b_{2})^2}{a_3}.\]
If not all the $a_i$ are equal then the Hamiltonian can be represented in
the form
\[\begin{array}{l}
H=a_{1} M_{1}^{2}+a_{2} M_{2}^{2}+a_{3} M_{3}^{2}+2 a_{2}a_{3}
M_{1}\gamma_{1}
+2 a_{3}a_{1}M_{2}\gamma_{2}+2 a_{1}a_{2} M_{3}\gamma_{3}\\[3mm]
\qquad +a_{1}(a_{2}-a_{3})^{2}\gamma_{1}^{2}
+a_{2}(a_{3}-a_{1})^{2}\gamma_{2}^{2}+a_{3}(a_{1}-a_{2})^{2}\gamma_{3}^{2}.
\end{array}\]
The additional integral is quadratic:
\[\begin{array}{l}
\!I=M_{1}^{2}+ M_{2}^{2}+M_{3}^{2}+2 (a_{2}+a_{3}) M_{1}\gamma_{1}
+2 (a_{3}+a_{1}) M_{2}\gamma_{2}+2 (a_{1}+a_{2}) M_{3}\gamma_{3}\\[3mm]
\qquad +(a_{2}-a_{3})^{2}\gamma_{1}^{2}
+(a_{3}-a_{1})^{2}\gamma_{2}^{2}+(a_{1}-a_{2})^{2}\gamma_{3}^{2}.
\end{array}
\]
If $a_{1}=a_{2}=a_{3}$ we have to interchange the Hamiltonian $H$ and the
integral $I$ just as in the Clebsch case.

Recently in the paper \cite{sok1} by Sokolov a Hamiltonian with non-diagonal
matrix $B$ having an integral of fourth degree was presented.
One of the possible forms of this Hamiltonian is:
\begin{equation} \nonumber % \label{newHAM}
H=M_{1}^2 + M_{2}^2 + 2\, M_{3}^2 + 2\, (a_{1} \gamma_{1} +
a_{2} \gamma_{2})\, M_{3} - (a_{1}^2 + a_{2}^2)\,\gamma_{3}^2.
\end{equation}
The integral can be represented as a product
$I=k_{1}\, k_{2}$, where $k_{1}=M_{3}$ and
\[\begin{array}{l}
k_{2}=(M_{1}^2 + M_{2}^2+ M_{2}^2)\, M_{3} +
2\,(a_{1} M_{1} + a_{2} M_{2})\,(M_{1} \gamma_{1} + M_{2} \gamma_{2})\\[3mm]
\qquad +2\,(a_{1}\gamma_{1}+a_{2}\gamma_{2})\,M_{3}^{2}
    +(a_{1}\gamma_{1}+a_{2}\gamma_{2})^{2}\,M_{3} \\[3mm] \qquad -
    (a_{1}^2+a_{2}^2)\,(2 M_{1} \gamma_{1}+2
M_{2}\gamma_{2}+M_{3}\gamma_{3})
    \,\gamma_{3}.
\end{array}
\]
This case appears to be similar in its properties to the Kowalewski
case given below.

{\bf The problem of motion of a rigid body around a fixed point.}
Hamiltonians describing such situations have the form (\ref{diagHam}), where
$b_{i}=c_{i}=p_{i}=0$.
The following integrable cases are known.

{\it The Lagrange case:}
\begin{equation} \label{lagrange}
H=M_{1}^{2}+M_{2}^{2}+a_{3} M_{3}^{2}+q_{3} \gamma_{3}
\end{equation}
where $a_{3}, q_{3}$ are arbitrary.
The additional integral $I$ is linear: $I=M_{3}$.

{\it The Euler case:}
\begin{equation} \label{euler}
H=a_{1} M_{1}^{2}+a_{2} M_{2}^{2}+a_{3} M_{3}^{2}
\end{equation}
with an additional quadratic integral:
\[I= M_{1}^{2}+M_{2}^{2}+M_{3}^{2}.\]

{\it The Kowalewski case:}
\begin{equation} \label{koval}
H=M_{1}^{2}+M_{2}^{2}+2 M_{3}^{2}+q_{1} \gamma_{1}+q_{2} \gamma_{2}
\end{equation}
with arbitrary parameters $q_{1}, q_{2}$. The additional integral of
fourth degree can be written as $I=G_{1}^{2}+G_{2}^{2},$ where
\[G_{1}=M_{1}^{2}-M_{2}^{2}-q_{1}\gamma_{1}+q_{2} \gamma_{2},\qquad
G_{2}=2M_{1} M_{2}-q_{2}\gamma_{1} - q_{1} \gamma_{2}.\]

We note, that if $q_{2}\ne i q_{1},$ then the transformation
$M\rightarrow T\,M$, $\Gamma\rightarrow T\,\Gamma$, where
\begin{equation} \label{TT}
T=\pmatrix{\cos{\phi}&\sin{\phi}&0 \cr
-\sin{\phi}&\cos{\phi}&0 \cr
0&0&1\cr}
\end{equation}
can be used to make $q_{2}$ to zero.

{\bf Generalizations.} Terms linear in the moments $M_{i}$ that occur
in the Hamiltonian could be interpreted as an action of hydrostatic
forces (\cite{boma2}). We have no comments on the physical meaning of other
mixed Hamiltonians (\ref{HAMGEN}) (i.e. Hamiltonians having both $B\ne
0$ or $C\ne 0$, and $Q\ne 0$ ).

For example, an obvious hybrid of Lagrange's and Kirchhoff's Hamiltonians
(with additional hydrostatic member) is
\begin{equation}
\begin{array}{l}
H= M_{1}^2 + M_{2}^2 + s_{1} M_{3}^2 +
 s_{2}\gamma_{3} M_{3} + s_{3} \gamma_{3}^2 + s_{4} M_{3}+ s_{5}\gamma_{3},
\end{array}  \label{Ham1-1}
\end{equation}

\section{Results}  \label{results}%=======================================
\subsection{The classical case.}  \label{e3c2}
For the Hamiltonian (\ref{HAMGEN}) on $e(3)$ we consider the case of
$A$ being diagonal:
\[A= \mbox{diag}(a_{1},a_{2},a_{3}),\]
where
\[a_{1}=a_{2}\ne a_{3}, \qquad a_{i}\ne 0, \qquad i=1,2,3.\]
Without loss of generality matrix $B$ can be considered as low-triangular,
i.e.
\[b_{12}=b_{13}=b_{23}=0.\]

Note, that the addition of Casimir functions (\ref{cas}) to the Hamiltonian
does not
influence the equation of motion.
By subtracting an appropriate linear combination of $J_{1}$
and $J_{2}$ we can make $b_{11}=c_{11}=0$ and get
\begin{eqnarray}
H&=& M_{1}^2 + M_{2}^2 + a_{3} M_{3}^2 +  \nonumber \\
 & & b_{21} \gamma_{1} M_{2} + b_{31} \gamma_{1} M_{3} +
     b_{32} \gamma_{2} M_{3} +b_{22} \gamma_{2} M_{2} +
     b_{33} \gamma_{3} M_{3} +  \nonumber \\
 & & c_{12}\gamma_{1}\gamma_{2} + c_{13}\gamma_{1}\gamma_{3} +
     c_{22}\gamma_{2}^{2} + c_{23}\gamma_{2}\gamma_{3} +
     c_{33}\gamma_{3}^{2}+\label{Hamc}\\
 & & p_{1} M_{1}+  p_{2} M_{2} +  p_{3} M_{3} + q_{1}\gamma_{1} +
     q_{2}\gamma_{2}+ q_{3}\gamma_{3}. \nonumber
\end{eqnarray}

{\bf Lemma}. For a Hamiltonian of type (\ref{Hamc}), a canonical
transformation can be used to obtain $b_{21}=0.$

Thus we will consider only the following Hamiltonian:
\begin{eqnarray}
H&=& M_{1}^2 + M_{2}^2 + a_{3} M_{3}^2 +
  b_{31} \gamma_{1} M_{3} +  b_{32} \gamma_{2} M_{3}
+b_{22} \gamma_{2} M_{2} + b_{33} \gamma_{3} M_{3} + \nonumber \\
 & &  c_{12}\gamma_{1}\gamma_{2} + c_{13}\gamma_{1}\gamma_{3} +
c_{22}\gamma_{2}^{2} + c_{23}\gamma_{2}\gamma_{3} +
c_{33}\gamma_{3}^{2}+ \label{Hamob} \\
 & &  p_{1} M_{1}+  p_{2} M_{2} +  p_{3} M_{3} + q_{1}\gamma_{1} +
q_{2}\gamma_{2}+ q_{3}\gamma_{3}.\nonumber
\end{eqnarray}

{\bf Theorem 1}.{\it The Hamiltonian of the above form
commutes with a polynomial integral of $4^{\mbox{\scriptsize th}}$ degree
iff it coincides with one of the following}:
\begin{itemize}
\item (\ref{Ham1-1}),
{\it where}  $s_{i}$ -  {\it are arbitrary, or }
\item
\begin{equation}
\begin{array}{l}
H= M_{1}^2 + M_{2}^2 + M_{3}^2 + 2 s_{1}\gamma_{3} M_{3} - s_{1}^{2}
\gamma_{3}^2 + s_{2}\gamma_{1} + s_{3}\gamma_{2} + s_{4}\gamma_{3} +\\[4mm]
\qquad  \lambda \,\Big( 2 s_{1} M_{3}^2 + s_{2} M_{1}+  s_{3} M_{2} + s_{4}
M_{3} + s_{1}(s_{2}\gamma_{1} + s_{3}\gamma_{2})\Big),
\end{array}  \label{Ham2-1}
\end{equation}
\item
\begin{equation}
\begin{array}{l}
H= M_{1}^2 + M_{2}^2 + 2 M_{3}^2 + s_{1}( i \gamma_{1} + \gamma_{2})
M_{3}+\\[4mm]
\qquad s_{2} (-i M_{1} + M_{2}) + s_{3} M_{3} + s_{4}(i \gamma_{1} +
\gamma_{2})- s_{1}s_{2}\gamma_{3},
\end{array}  \label{Ham3-1}
\end{equation}
\item
\begin{equation}
\begin{array}{l}
H=M_{1}^2 + M_{2}^2 + 2 M_{3}^2 + 2(s_{1} \gamma_{1} + s_{2} \gamma_{2})\,
M_{3} - (s_{1}^2 +s_{2}^2) \gamma_{3}^2 +\\[4mm]
\qquad s_{3}\, (M_{3}+s_{1} \gamma_{1}+s_{2} \gamma_{2})+
s_{4} \gamma_{1}+s_{5} \gamma_{2},
\end{array}  \label{Ham4-1}
\end{equation}
{\it where} $s_{i}$ {\it are parameters constrained only by the
condition}:
\[s_{2} s_{5}+s_{1} s_{4}=0.\]
\end{itemize}

{\bf Remarks}\\
1. The case (\ref{Ham2-1}) was founded by Rubanovskiyi \cite{rub}. \\
2. Under the conditions $s_{1}= s_{2}=s_{3}=0$ the formula
(\ref{Ham4-1}) describes the Kowalewski case. If $s_{3}=
s_{4}=s_{5}=0$ we have the Sokolov case for which the general formula
(\ref{Ham4-1}) was presented in \cite{sok2} and the Lax pair
was found by Sokolov and A.V.\ Tsiganov \cite{soktsig1}. \\
3. The Hamiltonian (\ref{Ham3-1}) is probably new.

\subsection{Classical case with additional condition.}\label{e3c2null}
We now consider the case of the area integral $J_{2}$ being equal to zero.
In this case the following additional integrable Hamiltonians are known.

The Goryachev-Chaplygin Hamiltonian {\cite{gor,chap}}
\begin{equation}
\begin{array}{l}
H= M_{1}^2 + M_{2}^2 + 4 M_{3}^2 - s_{1}\gamma_{1}
\end{array}  \label{gor-chap}
\end{equation}
belongs to the class of Hamiltonians describing the
motion of a rigid body around a fixed point.
The additional integral is of third degree.

For the  Kirchhoff problem is known the Hamiltonian by Chaplygin:
\begin{equation}
\begin{array}{l}
H= M_{1}^2 + M_{2}^2 + 2 M_{3}^2 + c \gamma_{1}^{2}-
c \gamma_{2}^{2} .
\end{array}  \label{chap}
\end{equation}

Here the additional integral has fourth degree.

{\bf Theorem 2}. {\it The Hamiltonian of the form } (\ref{Hamob}) {\it
commutes with a polynomial integral of fourth degree with additional
condition} $(M,\gamma)=0$,  {\it iff it coincides with one of the
Hamiltonian from Theorem 1 or with the following}:
\begin{itemize}
\item
\begin{equation}
\begin{array}{l}
H= M_{1}^2 + M_{2}^2 + 2 M_{3}^2 +
 s_{1}( \gamma_{1}^{2} - \gamma_{2}^2)+ s_{2} \gamma_{1} \gamma_{2} +
 s_{3} M_{3}+s_{4}\gamma_{1}+ s_{5}\gamma_{2},
\end{array}  \label{HamN1}
\end{equation}
 {\it where}  $s_{i}$ -  {\it are arbitrary parameters}.
\end{itemize}

The Hamiltonian (\ref{HamN1}) is the generalization of
Chaplygin's case {\cite{gor,chap,yeh2}}.

\section{Quantum case.} \label{quantum}%==============
In the papers by Komarov, Sklyanin {\cite{kom1,kom2,sklyan} and others
quantum generalizations of classical Hamiltonians on $e(3)$ and
$so(4)$ were considered.  Similarly, instead of the Poisson bracket
(\ref{puas}) we investigate the commutator relations (\ref{com})
in an associative algebra with generators
$M_{1}, M_{2}, M_{3}$, $\gamma_{1},\gamma_{2},\gamma_{3}$.
We can consider $M_{i},\gamma_{i}$ to be operators. Due to commutator
relations
(\ref{com}) any monomial always could be ordered such, that $M_{i}$ come
before any $\gamma_{i}$
and indices increase (such a presentation is unique). Thus Hamiltonians
have the same form (\ref{HAMGEN}) as in the classical case, but the
multiplication is now non-commutative. In this case, any element $I$ from
the associative algebra that satisfies $[H,I]=0$ is called an integral.

{\bf Theorem 3}. {\it The Hamiltonian of the above form
commutes with a polynomial integral of $4^{\mbox{\scriptsize th}}$ degree
iff it coincides with one of the following}:

\begin{itemize}
\item
\begin{equation}
\begin{array}{l}
H= M_{1}^2 + M_{2}^2 + s_{1} M_{3}^2 +
 s_{2}\gamma_{3} M_{3} + s_{3} \gamma_{3}^2 + s_{4} M_{3}+ s_{5}\gamma_{3},
\end{array}  \label{Ham1-3}
\end{equation}
 {\it where}  $s_{i}$ -  {\it are arbitrary parameters};
\item
\begin{equation}
\begin{array}{l}
H= M_{1}^2 + M_{2}^2 + M_{3}^2 + 2 s_{1}\gamma_{3} M_{3} - s_{1}^{2}
\gamma_{3}^2 + s_{2}\gamma_{1} + s_{3}\gamma_{2} + s_{4}\gamma_{3} +\\[4mm]
\qquad  \lambda \,\Big( 2 s_{1} M_{3}^2 + s_{2} M_{1}+  s_{3} M_{2} + s_{4}
M_{3} + s_{1}(s_{2}\gamma_{1} + s_{3}\gamma_{2})\Big) ,
\end{array}  \label{Ham2-3}
\end{equation}
\item
\begin{equation}
\begin{array}{l}
H= M_{1}^2 + M_{2}^2 + 2 M_{3}^2 + s_{1}( i \gamma_{1} + \gamma_{2})
M_{3}+\\[4mm]
\qquad s_{2} (-i M_{1} + M_{2}) + s_{3} M_{3} + s_{4}(i \gamma_{1} +
\gamma_{2})- s_{1}s_{2}\gamma_{3},
\end{array}  \label{Ham3-3}
\end{equation}
\item
\begin{equation}
\begin{array}{l}
H=M_{1}^2 + M_{2}^2 + 2 M_{3}^2 + 2(s_{1} \gamma_{1} + s_{2} \gamma_{2})\,
M_{3} - (s_{1}^2 +s_{2}^2) \gamma_{3}^2 +\\[4mm]
\qquad s_{3}\, (M_{3}+s_{1} \gamma_{1}+s_{2} \gamma_{2})+
s_{4} \gamma_{1}+s_{5} \gamma_{2},
\end{array}  \label{Ham4-3}
\end{equation}
{\it where} $s_{i}$ {\it are parameters constrained only by the
condition}:
\[s_{2} s_{5}+s_{1} s_{4}=0;\]
\end{itemize}

Notice that this list looks exactly as the list of subsection \ref{e3c2}
although {\it a priori} this coincidence is not predictable.
The most interesting here is Hamiltonian (\ref{Ham4-3}), which is a quantum
generalization of the Sokolov Hamiltonian. If $s_{1}=s_{2}=s_{3}=0$ we have
the quantum Kowalewski Hamiltonian \cite{kom1,lap}.

\section*{Acknowledgements}
The authors are grateful to  V.V.~Sokolov for useful
discussions. The research was partially supported by RFBR grant 02-01-06670.

\end{document}